\documentclass[twocolumn,twoside,slac_two]{revtex4}
\usepackage{graphicx}
\usepackage{fancyhdr}
\def\dbar{\overline{D}{}^{\,0}}

\def\Db{\overline{D}{}^{\,0}}
\def\Dz{D^0}
\def\cp{$CP$}
\def\cpv{$CPV$}
\def\ra{\!\rightarrow\!}

\def\dkpi{$D^0\ra K^+\pi^-$}
\def\dkk{$D^0\ra K^+K^-$}
\def\dpipi{$D^0\ra\pi^+\pi^-$}

\def\dkspp{$D^0\ra K^0_S\,\pi^+\pi^-$}
\def\ycp{$y^{}_{\rm CP}$}

\def\mevm{~MeV/$c^2$}

\def\babar{{\it BABAR\/}}

\begin{document}

\title{\boldmath $D^0$ Mixing at Belle }

\author{L.~M.~Zhang (on
behalf of the Belle Collaboration)} \affiliation{Syracuse
University, Syracuse, New York 13244, USA}

\begin{abstract}
We report the recent two results of $D^0$-$\dbar$ mixing studies at
Belle in $D^0\to K^+K^-/\pi^+\pi^-$ and $D^0\to K_S^0\pi^+\pi^-$
decays. The former measures the relative difference of the lifetimes
$y_{CP}$, giving the evidence of $D^0$-$\dbar$ mixing; the latter
measures the $\Dz$ mixing parameters $x$ and $y$.
\end{abstract}

\maketitle

\thispagestyle{fancy}

\section{Introduction}

Mixing phenomenon, i.e. the oscillation of a neutral meson into its
corresponding anti-meson as a function of time, has been observed in
the $K^0$, $B^0$, and most recently $B^0_s$ systems. This process is
also possible in the $D$-meson system, but has not previously been
observed.

Mixing in heavy flavor systems such as that of $B^0$ and $B_s^0$ is
governed by the short-distance box diagram. However, in the $D^0$
system this diagram is both GIM-suppressed and
doubly-Cabibbo-suppressed relative to the amplitude dominating the
decay width, and thus the short-distance rate is very small.
Consequently, $D^0$-$\Db$ mixing is expected to be dominated by
long-distance processes that are difficult to calculate; theoretical
estimates for the mixing parameters $x=(m_1-m_2)/\Gamma$ and
$y=(\Gamma_1-\Gamma_2)/2\Gamma$ range over two-three orders of
magnitude \cite{Petrov}. Here, $m_1, m_2$ ($\Gamma_1,\Gamma_2$) are
the masses (decay widths) of the mass eigenstates
$|D_{1,2}\rangle=p|\Dz\rangle\pm q|\Db\rangle$, and
$\Gamma=(\Gamma_1+\Gamma_2)/2$.
The parameters $p$ and $q$ are complex coefficients satisfying
$|p|^2+|q|^2=1$.

The general experimental method identifies the flavor of the neutral
$D$ meson when produced by reconstructing the decay $D^{*+}\to
D^0\pi^+$ or $D^{*-}\to \Db \pi^-$ \cite{charge_conjugates}; the
charge of the accompanying pion identifies the $D$ flavor. Because
the energy release in $D^{*}$ decays is only  $\sim6$ MeV, the
background is largely suppressed. The $\Dz$ decay time ($t$) is
calculated via $(l/p)\times m_{\Dz}$, where $l$ is the distance
between the $D^*$ and $\Dz$ decay vertices and $p$ is the $\Dz$
momentum. The $D^*$ vertex position is taken to be the intersection
of the $\Dz$ momentum with the beamspot profile. To reject $D^{(*)}$
decays originating from $B$ decays, one requires $p_{D^*}>2.5$ GeV,
which is the kinematic endpoint.

\section{{\boldmath \cp-eigenstates $K^+K^-$ and $\pi^+\pi^-$}}
We have studied the decays to $CP$ eigenstates \dkk and \dpipi;
treating the decay-time distributions as exponential, we measured
the quantity
\begin{equation}
y_{CP}=\frac{\tau_{K^-\pi^+}}{\tau_{K^+K^-}}-1,
\end{equation}
where $\tau_{K^-\pi^+}$ and $\tau_{K^+K^-}$ are the lifetimes of
$\Dz \to K^-\pi^+$ and \dkk (or \dpipi) decays. It can be shown that
$y_{CP}=y\cos\phi-\frac{1}{2}A_Mx\sin\phi$~\cite{ycpeqn}, where
$A_M$ parameterizes $CPV$ in mixing and $\phi$ is a weak phase. If
$CP$ is conserved, $A_M=\phi=0$ and $y_{CP}=y$.  This method has
been used by numerous experiments to constrain
\ycp~\cite{ycp_references}. Our measurement, based on 540~fb$^{-1}$
data, yields a nonzero value of $y_{CP}$ with $>3\sigma$
significance~\cite{belle_kk}. We also searched for $CPV$ by
measuring the quantity
\begin{equation}
A_{\Gamma}=\frac{\tau(\dbar\to K^-K^+)-\tau(D^0\to
K^+K^-)}{\tau(\dbar\to K^-K^+)+\tau(D^0\ra K^+K^-)};
\end{equation}
this observable equals
$A_{\Gamma}=\frac{1}{2}A_My\cos\phi-x\sin\phi$~\cite{ycpeqn}.

We reconstruct $D^{*+}\to D^0\pi_s^+$ decays and \dkk, $K^-\pi^+$,
and $\pi^+\pi^-$. Candidate $\Dz$ mesons are selected using two
kinematic observables: the invariant mass of the $\Dz$ decay
products, $M$, and the energy release in the $D^{*+}$ decay,
$Q=(M_{D^*}-M-m_{\pi^+})c^2$. According to Monte Carlo (MC)
simulated distributions of $t$, $M$ and $Q$, background events fall
into four categories: (1) combinatorial, with zero apparent
lifetime; (2) true $D^0$ mesons combined with random slow pions
(this has the same apparent lifetime as the signal) (3) $D^0$ decays
to three or more particles, and (4) other charm hadron decays. The
apparent lifetime of the latter two categories is 10-30\% larger
than $\tau_{D^0}$.

For the lifetime measurements, we select the events satisfying
$|\Delta M|/\sigma_M<2.3$, $|Q-5.9$ MeV$|<0.80$ MeV and
$\sigma_t<370$ fs, where $\Delta M\equiv M-m_{D^0}$, and $\sigma_t$
is the decay time uncertainties calculated event-by-event. The
invariant mass resolution $\sigma_M$ varies from 5.5-6.8 MeV/$c^2$,
depending on the decay channel. The selection criteria are chosen to
minimize the expected statistical error on $y_{CP}$ using the MC. We
find $111\times 10^3$ $K^+K^-$, $1.22\times10^6$ $K^-\pi^+$ and
$49\times 10^3$ $\pi^+\pi^-$ signal events, with purities of 98\%,
99\% and 92\% respectively.

The relative lifetime difference $y_{CP}$ is determined by
performing a simultaneous binned maximum likelihood fit to the \dkk,
\dkpi, \dpipi decay time distributions. Each distribution is assumed
to be a sum of signal and background contributions, with the signal
contribution being a convolution of an exponential and a detector
resolution function,
\begin{equation}
dN/dt = \frac{N_{\rm sig}}{\tau}\int e^{-t^\prime}{\tau} \,\cdot\,
R(t-t^\prime)\,dt^\prime+B(t).
\end{equation}

The resolution function $R(t-t^{\prime})$ is constructed from the
normalized distribution of the decay time uncertainties $\sigma_t$.
The $\sigma_t$ of a reconstructed event ideally represents an
uncertainty with a Gaussian probability density: in this case, bin
$i$ in the $\sigma_t$ distribution is taken to correspond to a
Gaussian resolution term of width $\sigma_i$, with a weight given by
the fraction $f_i$ of events in that bin. However, the distribution
of ``pulls", i.e. the normalized residuals $(t_{\rm rec}-t_{\rm
gen})/\sigma_t$ (where $t_{\rm rec}$ and $t_{\rm gen}$ are
reconstructed and generated decay times), is not well-described by a
Gaussian. We found that this distribution can be fitted with a sum
of three Gaussians of different widths $\sigma_k^{\rm pull}$ and
fractions $w_k$, constrained to the same mean. Therefore, we choose
the parameterization
\begin{equation}\label{res}
R(t-t^{\prime})=\sum_{i=1}^n f_i \sum_{i=1}^3 w_k G(t-t^{\prime};
\sigma_{ik},t_0),
\end{equation}
with $\sigma_{ik}=s_k\sigma_k^{\rm pull}\sigma_i$, where the $s_k$
are three scale factors introduced to account for differences
between the simulated and real $\sigma_k^{\rm pull}$, and $t_0$
allows for a (common) offset of the Gaussian terms from zero.

The background $B(t)$ is parameterized assuming two lifetime
components: an exponential and a $\delta$ function, each convolved
with corresponding resolution functions as parameterized by Eq.
(\ref{res}). Separate $B(t)$ parameters for each final state are
determined by fits to the $t$ distributions of events in $M$
sidebands. The MC is used to select the sideband region that best
reproduces the timing distribution of background events in the
signal region.

\begin{figure}[!hbtp]
\centering
\includegraphics[width=80mm]{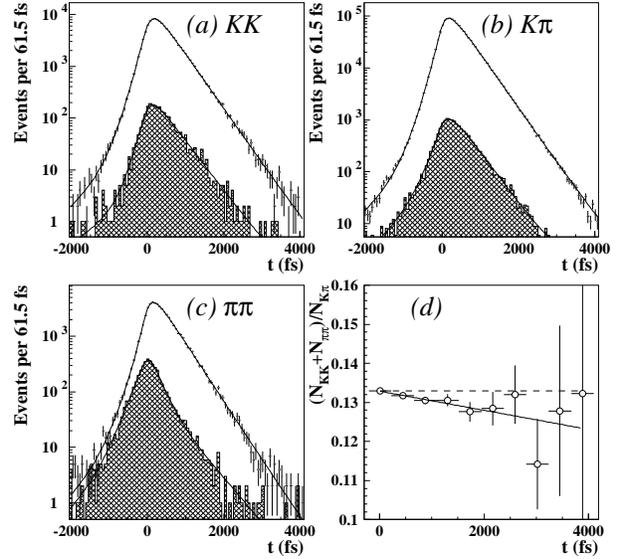}
\caption{Projections of the decay-time fit superimposed on the data
for \dkk, $D^0\ra K^-\pi^+$, and \dpipi\ decays. The hatched area
represents the background contribution. Plot {\it (d)} shows the
ratio of decay-time distributions for $D^0\ra(K^+K^- +\pi^+\pi^-)$
and $D^0\ra K^-\pi^+$; the solid line is a fit to the points.}
\label{fig:ycp_belle}
\end{figure}

Fitting the $K^-\pi^+$, $K^+K^-$, and $\pi^+\pi^-$ decay time
distributions (Figs.~\ref{fig:ycp_belle}(a)-(c)) shows a
statistically significant difference between the $K^-\pi^+$ and
$h^+h^-$ lifetimes. The effect is visible in
Fig.~\ref{fig:ycp_belle}d, which plots the ratio of event yields
$N^{}_{h^+h^-}/N^{}_{K\pi}$ as a function of decay time. The fitted
lifetime of $D^0$ meson in the $K^-\pi^+$ final states is
$408.7\,\pm 0.6$~fs, which is consistent with the PDG
value~\cite{pdg} (and actually has greater statistical precision).
We measure
\begin{eqnarray}
y^{}_{\rm CP} & = & (1.31\,\pm 0.32\,\pm 0.25)\%\,,
\end{eqnarray}
which deviates from zero by $3.2\sigma$. The systematic error is
dominated by uncertainty in the background decay time distribution,
variation of selection criteria, and the assumption that $t^{}_0$ is
equal for all three final states. The analysis also measures
\begin{eqnarray}
A^{}_\Gamma & = & (0.01\,\pm 0.30\,\pm 0.15)\%\,,
\end{eqnarray}
which is consistent with zero (no \cpv). The sources of
systematic error for $A^{}_\Gamma$ are similar to those for~\ycp.

\section{\boldmath Dalitz Plot Analysis of \dkspp}

The time dependence of the Dalitz plot for \dkspp\ decays
is sensitive to mixing parameters $x$ and $y$ without ambiguity
due to strong phases.
For a particular point in the Dalitz plot $(m^2_+,m^2_-)$, where
$m^{}_+\equiv m(K^0_S\,\pi^+)$ and $m^{}_-\equiv m(K^0_S\,\pi^-)$,
the overall decay amplitude is
\begin{eqnarray}
{\cal A}^{}_{D^0}(m^2_+,m^2_-)\,\frac{e^{}_1(t)+e^{}_2(t)}{2} & + & \nonumber \\
 & & \hskip-1.2in
\left(\frac{q}{p}\right)
{\cal A}^{}_{\dbar}(m^2_+,m^2_-)\,\frac{e^{}_1(t)-e^{}_2(t)}{2}\,,
\label{eqn:kspp_amplitudes}
\end{eqnarray}
where $e^{}_{(1,2)}(t)=e^{-(im^{}_{1,2}+\Gamma^{}_{1,2}/2)t}$.
The first term represents the (time-dependent) amplitude for
\dkspp, and the second term represents the amplitude
for $D^0\!\ra\!\dbar\!\ra\!K^0_S\,\pi^+\pi^-$. Taking the
modulus squared of Eq.~(\ref{eqn:kspp_amplitudes}) gives
the decay rate or, equivalently, the density of points
$\rho(m^2_+,m^2_-;t)$. The result contains terms proportional
to  $\cosh(y\,\Gamma t)$, $\cos(x\,\Gamma t)$, and
$\sin(x\,\Gamma t)$, and thus fitting the time-dependence
of $\rho(m^2_+,m^2_-;t)$ determines $x$ and~$y$.
This method was developed by CLEO~\cite{cleo_kspp}.

To use Eq.~(\ref{eqn:kspp_amplitudes}) requires
choosing a model for the decay amplitudes
${\cal A}^{}_{D^0,\dbar}(m^2_+,m^2_-)$. This is usually taken
to be the ``isobar model''~\cite{isobar}, and thus, in addition
to $x$ and $y$, one also fits for the magnitudes and phases of
various intermediate states. Specifically,
${\cal A}^{}_{D^0}(m^2_+,m^2_-)=\sum_j a^{}_j\,e^{i\delta^{}_j}\,A^{}_j$,
where $\delta^{}_j$ is a strong phase, $A^{}_j$ is the
product of a relativistic Breit-Wigner function and
Blatt-Weiskopf form factors, and the parameter $j$
runs over all intermediate states. This sum includes
possible scalar resonances and, typically, a constant
non-resonant term. For no direct \cpv,
${\cal A}^{}_{\dbar}(m^2_+,m^2_-)={\cal A}^{}_{D^0}(m^2_-,m^2_+)$;
otherwise, one must consider separate decay parameters
$(a^{}_j,\delta^{}_j)$ for $D^0$ decays and
$(\bar{a}^{}_j,\bar{\delta}^{}_j)$ for $\dbar$
decays.

We have fit a large \dkspp\ sample selected from 540~fb$^{-1}$ of
data~\cite{belle_kspp}. The analysis proceeds in two steps. First,
signal and background yields are determined from a two-dimensional
fit to variables $M(K\pi\pi)$ and $Q = M(\pi^{}_s
K\pi\pi)-M(K\pi\pi)-m_{\pi^+}$. Within a signal region
$|M(K\pi\pi)-m^{}_{D^0}|\!<\!15$\mevm\ and $|Q-5.9{\rm\
MeV}|\!<\!1.0$~MeV (corresponding to $3\sigma$ in resolution), there
are 534\,000 signal candidates with 95\% purity. These events are
fit for $x$ and $y$; the (unbinned ML) fit variables are $m^2_+$,
$m^2_-$, and the decay time~$t$. Most of the background is
combinatoric, i.e., the $D^0$ candidate results from a random
combination of tracks. The decay-time distribution of this
background is modeled as the sum of a delta function and an
exponential function convolved with a Gaussian resolution function,
and all parameters are determined from fitting events in the
sideband $30{\rm\ MeV}/c^2<|M(K\pi\pi)-m^{}_{D^0}|<55$\mevm.

\begin{figure}[!hbtp]
\centering
\includegraphics[width=39mm]{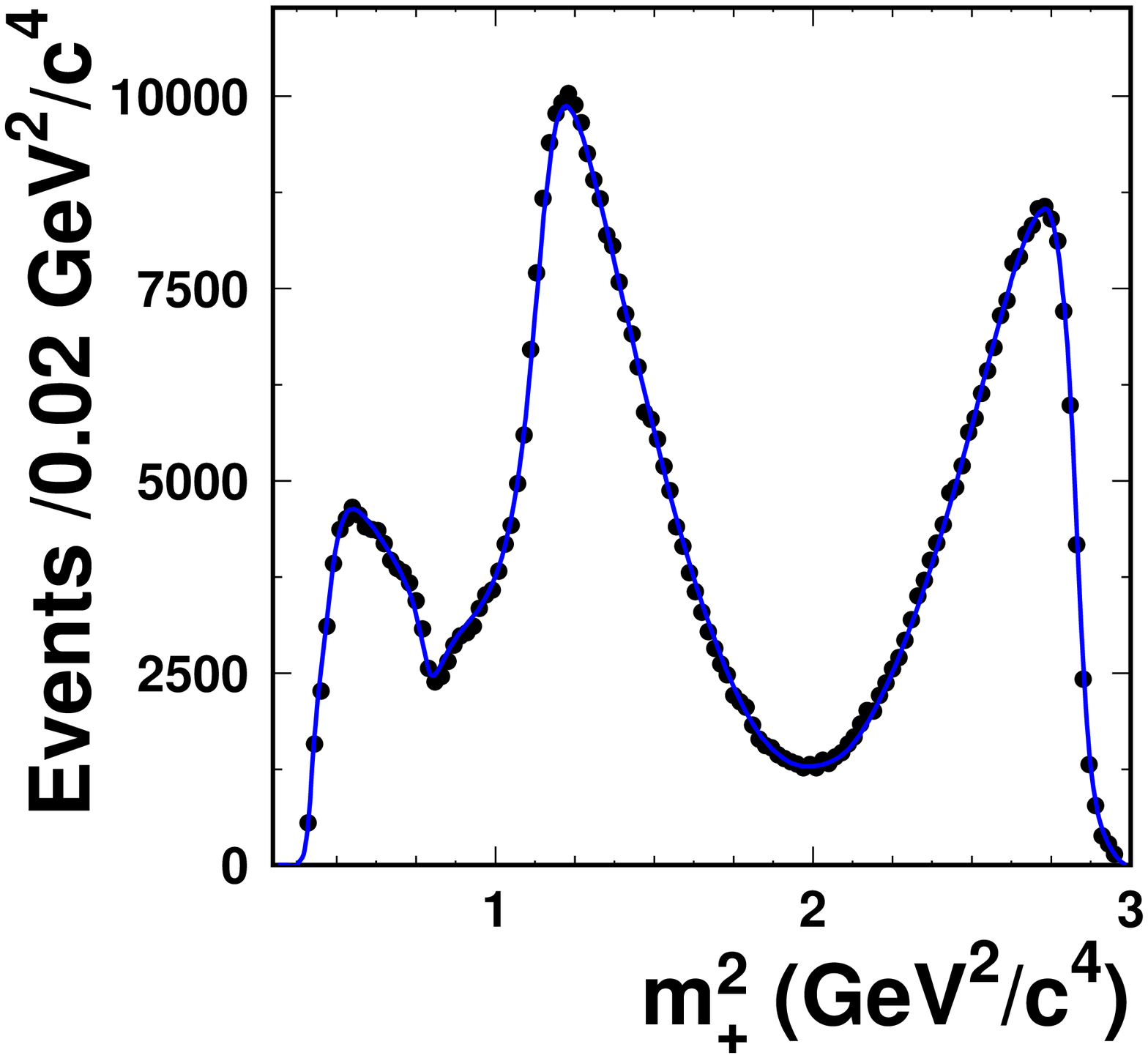}
\hskip0.05in
\includegraphics[width=39mm]{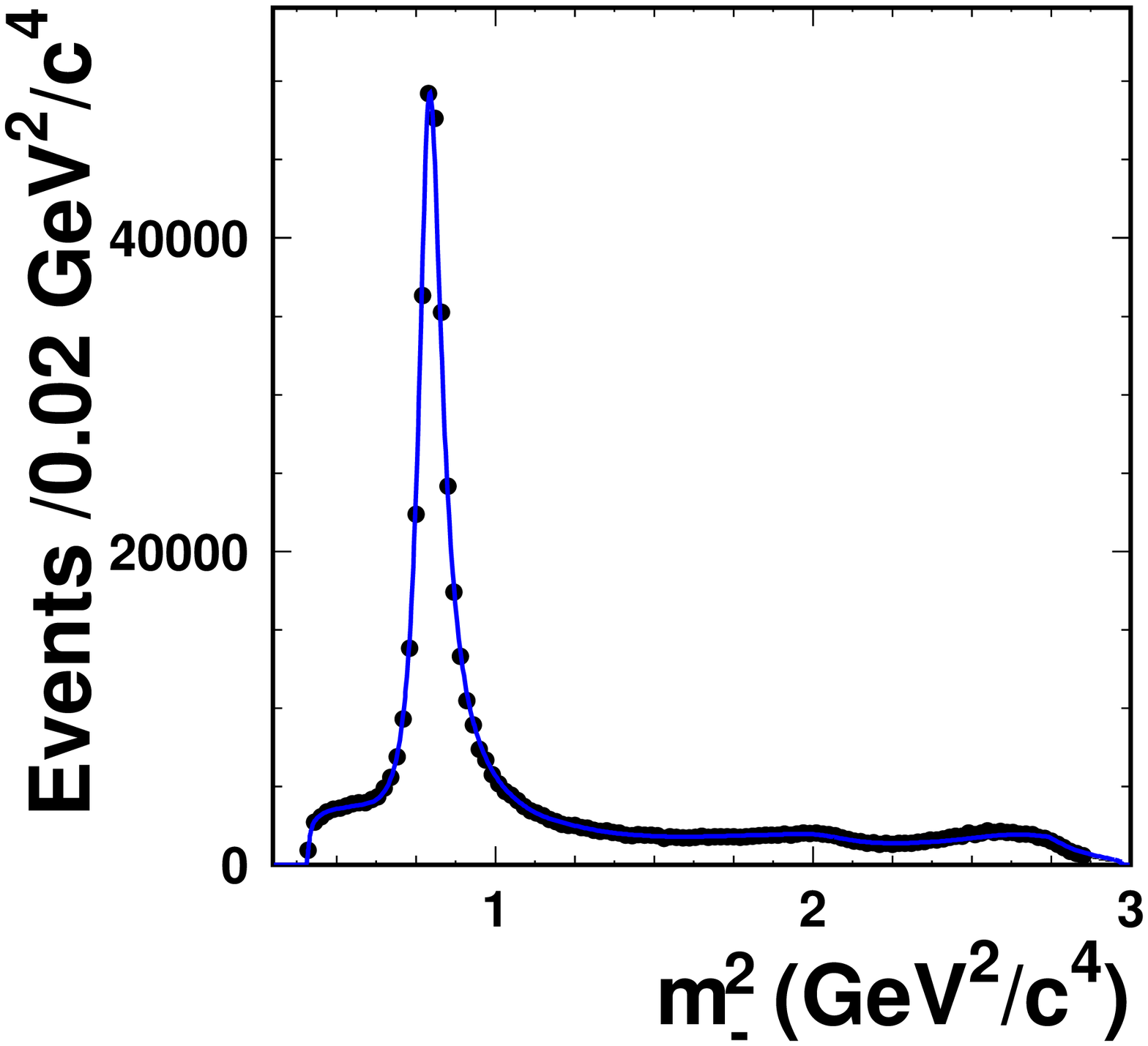}
\vskip0.10in
\includegraphics[width=39mm]{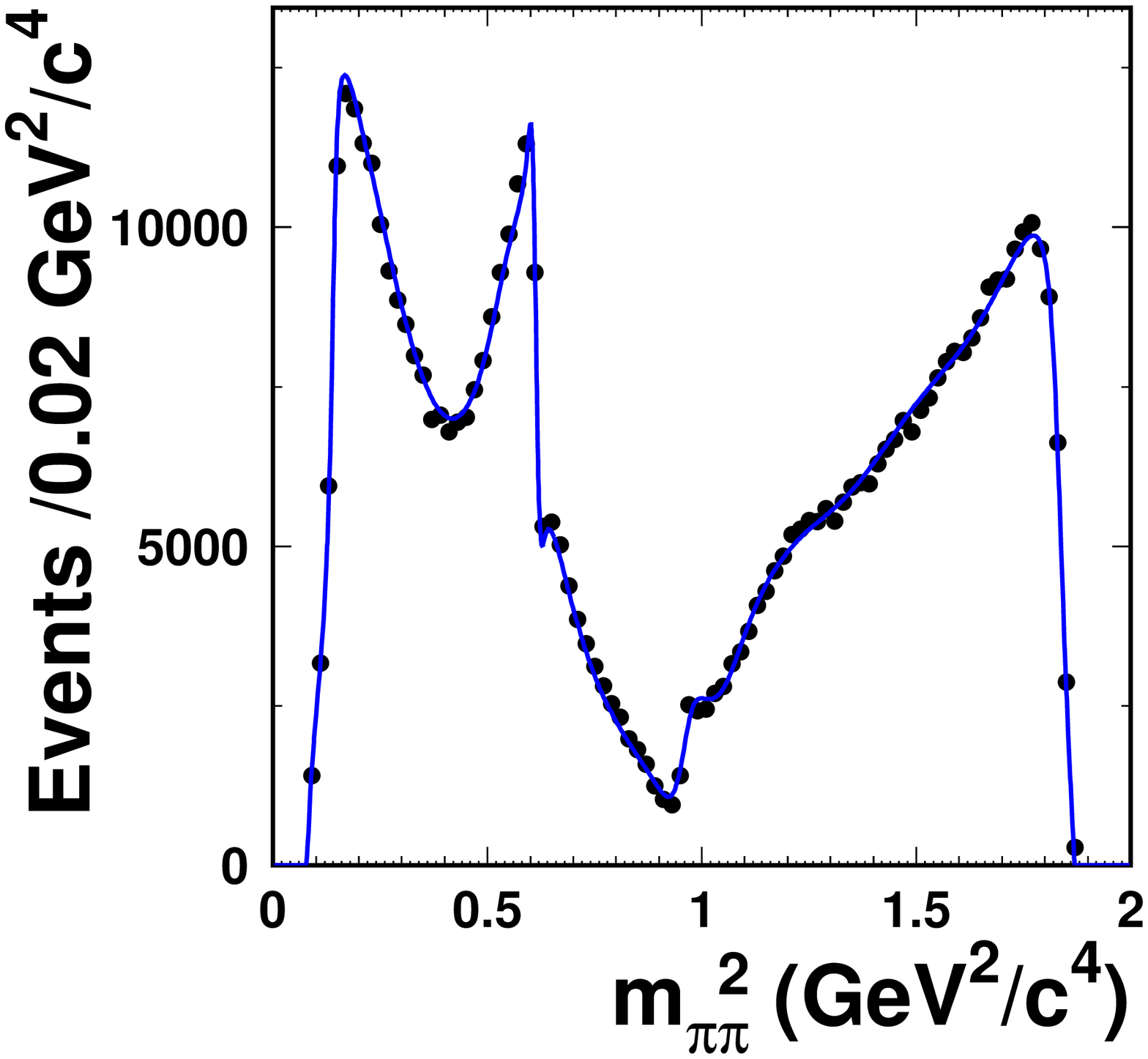}
\hskip0.05in
\includegraphics[width=39mm]{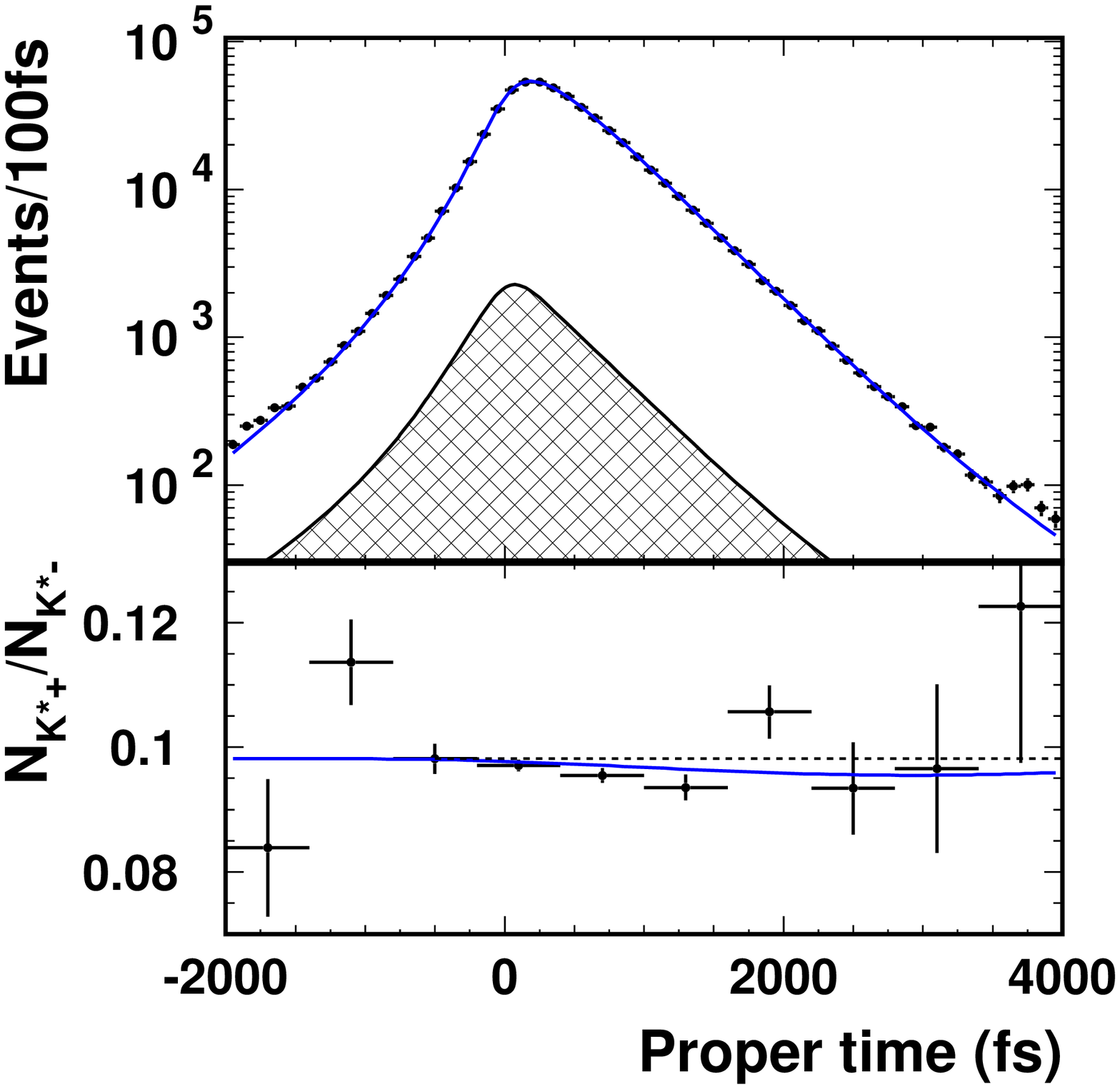}
\caption{Projection of the unbinned ML fit superimposed on the data
for \dkspp\ decays. In {\it (d)}, the hatched area represents the
combinatorial background contribution, and the lower plot shows the
ratio of decay-time distributions for events in the $K^*(892)^+$ and
$K^*(892)^-$ regions, where sensitivity to $(x,y)$ is highest.}
\label{fig:kspp_projections}
\end{figure}

The results from two separate fits are listed in
Table~\ref{tab:kspp_belle}. In the first fit \cp\
conservation is assumed, i.e., $q/p\!=\!1$ and
${\cal A}^{}_{\dbar}(m^2_+,m^2_-)={\cal A}^{}_{D^0}(m^2_-,m^2_+)$.
The free parameters are $x,y,\tau^{}_{D^0}$, some
timing resolution function parameters, and decay
model parameters $(a^{}_r,\delta^{}_r)$. The results
for the latter are listed in Table~\ref{tab:kspp_resonances}.
The results for $x$ and $y$ indicate that $x$ is
positive, about $2\sigma$ from zero. Projections
of the fit are shown in Fig.~\ref{fig:kspp_projections}.
The fit also yields $\tau^{}_D=(409.9\pm 1.0)$~fs, which is
consistent with the PDG value~\cite{pdg} (and actually has greater
statistical precision).

\begin{table}[!hbtp]
\renewcommand{\arraystretch}{1.3}
\caption{Fit results and 95\% C.L. intervals for $x$ and $y$, from
analysis of \dkspp\ decays. The errors are statistical, experimental
systematic, and decay-model systematic, respectively.}
\label{tab:kspp_belle}
\begin{tabular}{llcc}\hline\hline
Fit   & Param. & Result  & 95\% C.L. inter. \\
\hline
No    & $x\,(\%)$
  & $0.80\,\pm 0.29\,^{+0.09\,+0.10}_{-0.07\,-0.14}$ & $(0.0,1.6)$ \\
$CPV$ & $y\,(\%)$
  & $0.33\,\pm 0.24\,^{+0.08\,+0.06}_{-0.12\,-0.08}$ & $(-0.34,0.96)$ \\
\hline
$CPV$ & $x\,(\%)$
  & $0.81\,\pm0.30\,^{+0.10\,+0.09}_{-0.07\,-0.16}$ & $|x|<1.6$  \\
      & $y\,(\%)$
  & $0.37\,\pm0.25\,^{+0.07\,+0.07}_{-0.13\,-0.08}$ & $|y|<1.04$ \\
      & $|q/p|$ &$0.86\,^{+0.30\,+0.06}_{-0.29\,-0.03}\,\pm 0.08$ & $-$ \\
      & $\phi\,(^\circ)$ & $-14\,^{+16\,+5\,+2}_{-18\,-3\,-4}$ & $-$ \\
\hline\hline
\end{tabular}
\end{table}

\begin{table}[!hbtp]
\caption{Fit results for \dkspp\ Dalitz plot parameters. The errors
are statistical only. The fit fraction is defined as the ratio of
the integral $\int |a^{}_r\,{\cal
A}^{}_r(m^2_-\,,m^2_+)|^2\,dm^2_-\,dm^2_+$
to 
$\int |\sum_{r=1}^n a^{}_r\,e^{i\delta^{}_r}
     {\cal A}^{}_r(m^2_-\,,m^2_+)|^2\,dm^2_-\,dm^2_+$.}
\label{tab:kspp_resonances}
\begin{tabular}{lccc}\hline\hline
Resonance & Amplitude& Phase (deg) &Fit fraction \\
\hline
$K^*(892)^-$    & $1.629\pm0.006$  & $134.3\pm0.3$  & 0.6227 \\
$K_0^*(1430)^-$ & $2.12\pm0.02$    & $-0.9\pm0.8$   & 0.0724 \\
$K_2^*(1430)^-$ & $0.87\pm0.02$    & $-47.3\pm1.2$  & 0.0133 \\
$K^*(1410)^-$   & $0.65\pm0.03$    & $111\pm4$      & 0.0048 \\
$K^*(1680)^-$   & $0.60\pm0.25$    & $147\pm29 $     & 0.0002 \\
\hline
$K^*(892)^+$    & $0.152\pm0.003$  & $-37.5\pm1.3$  & 0.0054 \\
$K_0^*(1430)^+$ & $0.541\pm0.019$  & $ 91.8\pm2.1$  & 0.0047 \\
$K_2^*(1430)^+$ & $0.276\pm0.013$  & $-106\pm3$     & 0.0013 \\
$K^*(1410)^+$   & $0.33\pm0.02$  & $-102\pm4$     & 0.0013 \\
$K^*(1680)^+$   & $0.73\pm0.16$    & $103\pm11$      & 0.0004 \\
\hline
$\rho(770)$     & 1 (fixed)          & 0 (fixed)      & 0.2111 \\
$\omega(782)$   & $0.0380\pm0.0007$  & $115.1\pm1.1$  & 0.0063 \\
$f_0(980)$      & $0.380\pm0.004$    & $-147.1\pm1.1$ & 0.0452 \\
$f_0(1370)$     & $1.46\pm0.05$      & $98.6 \pm1.8$  & 0.0162 \\
$f_2(1270)$     & $1.43\pm0.02$      & $-13.6\pm1.2$  & 0.0180 \\
$\rho(1450)$    & $0.72\pm0.04$      & $41\pm7$   & 0.0024 \\
$\sigma_1$      & $1.39\pm0.02$    & $-146.6\pm0.9$     & 0.0914 \\
$\sigma_2$      & $0.267\pm0.013$    & $-157\pm3 $    & 0.0088 \\
\hline
 NR             & $2.36\pm0.07$      & $155\pm2$      & 0.0615 \\ \hline\hline
\end{tabular}
\end{table}

For the second fit, \cpv\ is allowed and the $D^0$ and $\dbar$
samples are considered separately. This introduces additional
parameters $|q/p|$, ${\rm Arg}(q/p)=\phi$, and
$(\bar{a}^{}_j,\bar{\delta}^{}_j)$. The fit gives two equivalent
solutions, $(x,y,\phi)$ and $(-x,-y,\phi+\pi)$. Aside from this
possible sign change, the effect upon $x$ and $y$ is small, and the
results for $|q/p|$ and $\phi$ are consistent with no~\cpv. The sets
of Dalitz parameters $(a^{}_r,\delta^{}_r)$ and
$(\bar{a}^{}_r,\bar{\delta}^{}_r)$ are consistent with each other,
indicating no direct~\cpv. Taking $a^{}_j\!=\!\bar{a}^{}_j$ and
$\delta^{}_j\!=\!\bar{\delta}^{}_j$ (i.e., no direct \cpv) and
repeating the fit gives $|q/p|=0.95\,^{+0.22}_{-0.20}$ and $\phi =
(-2\,^{+10}_{-11})^\circ$.

The dominant systematic errors are from the time dependence of the
Dalitz plot background, and the effect of the $p^{}_{D^*}$ momentum
cut used to reject $D^*$'s originating from $B$ decays. The default
fit includes $\pi\pi$ scalar resonances $\sigma^{}_1$ and
$\sigma^{}_2$; when evaluating systematic errors, the fit is
repeated without any $\pi\pi$ scalar resonances using $K$-matrix
formalism~\cite{K-matrix}. The influence upon $x$ and $y$ is small
and included as a systematic error.

\begin{figure}[!hbtp]
\centering
\includegraphics[width=80mm]{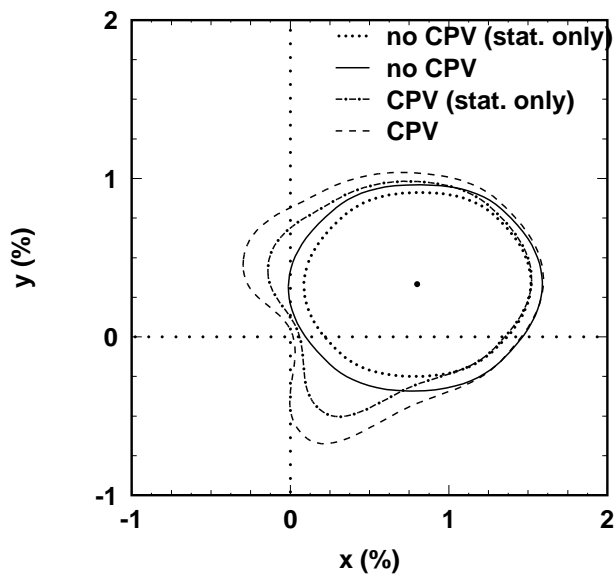}
\caption{95\% C.L. contours for $(x,y)$: dotted (solid) is
statistical (statistical plus systematic) contour for no \cpv;
dashed-dotted (dashed) is statistical (statistical plus systematic)
contour allowing for \cpv. The point is the best-fit value for
no~\cpv.} \label{fig:kspp_conf_regions}
\end{figure}

The 95\% C.L.\ contour for $(x,y)$ is plotted in
Fig.~\ref{fig:kspp_conf_regions}. The contour is obtained from the
locus of points where $-2\ln{\cal L}$ rises by 5.99 units from the
minimum value; the distance of the points from the origin is
subsequently rescaled to include systematic uncertainty. We note
that for the \cpv-allowed case, the reflections of the contours
through the origin are also allowed regions.

\begin{thebibliography}{99}


\bibitem{Petrov}
  A.\,A.\,Petrov, Int. J. Mod. Phys. A{\bf21}, 5686 (2006);
 arXiv:hep-ph/0611361.

\bibitem{charge_conjugates}
  Charge-conjugate modes are included unless noted otherwise.


\bibitem{ycpeqn}
S.~Bergmann {\it et al.}, Phys.\ Lett.~B~{\bf 486}, 418 (2000).


\bibitem{ycp_references}
  E.\ M.\ Aitala {\it et al.} (E791), Phys.\ Rev.\ Lett.~{\bf 83}, 32 (1999);
  J.~M.~Link {\it et al.} (FOCUS), Phys.\ Lett.~B~{\bf 485}, 62 (2000);
  S.~E.~Csorna {\it et al.} (CLEO), Phys.\ Rev.~D~{\bf 65}, 092001 (2002);
  B.~Aubert {\it et al.} (\babar), Phys.\ Rev.\ Lett.~{\bf 91}, 121801 (2003).

\bibitem{belle_kk}
  M.\ Staric {\it et al.} (Belle), Phys.\ Rev.\ Lett.~{\bf 98}, 211803 (2007).

\bibitem{pdg}
  W.-M. Yao {\it et al.} (PDG), Jour.\ of Phys.~G~{\bf 33}, 1 (2006).

\bibitem{cleo_kspp}
  D.\ M.\ Asner {\it et al.} (CLEO), Phys.\ Rev.~D~{\bf 72}, 012001 (2005);
  arXiv:hep-ex/0503045 (revised April, 2007).

\bibitem{isobar}
  A.\ Poluektov {\it et al.} (Belle), Phys.\ Rev.~D~{\bf 73}, 112009 (2006);
  S.~Kopp {\it et al.} (CLEO), Phys.\ Rev.~D~{\bf 63}, 092001 (2001).

\bibitem{belle_kspp}
  L.\ M.\ Zhang {\it et al.} (Belle), Phys.\ Rev.\ Lett.~{\bf 99}, 131803 (2007).

\bibitem{K-matrix}
  J.\ M.\ Link {\it et al.} (FOCUS), Phys.\ Lett.~B~{\bf 585}, 200 (2004);
  B.~Aubert {\it et al.} (\babar), arXiv:hep-ex/0507101 (2005).

\end{thebibliography}
\end{document}